\documentclass[nolinenumbers,twocolumn]{aastex631}
\usepackage{amsmath}
\usepackage{diagbox} 
\usepackage{gensymb}
\usepackage{diagbox}
\usepackage{multirow}
\usepackage{array}
\usepackage{appendix}
\usepackage{threeparttable}
\usepackage{comment}

\DeclareUnicodeCharacter{02BC}{<definition>}
\DeclareUnicodeCharacter{FF1A}{<definition>}

\begin{document}

\title{Joint constraint on the propagation origin of the cosmic-ray spectral knee from energy spectrum and anisotropy observations}

\correspondingauthor{Lin Nie}
\email{nielin@ihep.ac.cn}

\author{Hua Yue}
\affiliation{State Key Laboratory of Particle Astrophyics, Institute of High Energy Physics, Chinese Academy of Sciences, 100049 Beijing, China
}
\affiliation{School of Physical Sciences, University of Chinese Academy of Sciences, No.19(A) Yuquan Road, Beijing, China
}

\author{Lin Nie$^{\dag}$}
\affiliation{School of Physical Science and Technology, Southwest Jiaotong University, Chengdu, 610031, China}
\affiliation{State Key Laboratory of Particle Astrophyics, Institute of High Energy Physics, Chinese Academy of Sciences, 100049 Beijing, China
}

\author{Yiqing Guo}
\affiliation{State Key Laboratory of Particle Astrophyics, Institute of High Energy Physics, Chinese Academy of Sciences, 100049 Beijing, China
}
\affiliation{School of Physical Sciences, University of Chinese Academy of Sciences, No.19(A) Yuquan Road, Beijing, China
}
\affiliation{TIANFU Cosmic Ray Research Center, Chengdu, Sichuan, China}
\author{Hongbo Hu}
\affiliation{State Key Laboratory of Particle Astrophyics, Institute of High Energy Physics, Chinese Academy of Sciences, 100049 Beijing, China
}
\affiliation{School of Physical Sciences, University of Chinese Academy of Sciences, No.19(A) Yuquan Road, Beijing, China
}
\affiliation{TIANFU Cosmic Ray Research Center, Chengdu, Sichuan, China}
 
\begin{abstract}
The origin mechanism of the cosmic-ray knee region remains an unresolved mystery, with acceleration, interaction, and propagation models drawing significant attention. The latest experimental observations of the PeV total spectrum, composition energy spectrum, and anisotropy—particularly the precise measurements of the proton spectrum by the LHAASO experiment—have provided crucial breakthroughs in uncovering its origin. Based on the latest LHAASO measurements of the proton energy spectrum, combined with cosmic-ray spectral and anisotropy data, this study proposes that the spectral index variation in the knee region arises from changes in the propagation coefficient. By introducing a knee position $\rm \mathcal{R}_{knee}$ and an index variation $\rm \delta_{knee}$, we construct a rigidity-dependent double-power-law diffusion model to reproduce the knee-region spectral structure. Through modifications to the diffusion coefficient, we successfully replicate the observed knee-region spectral structure in the LHAASO proton spectrum and calculate the corresponding anisotropy. Under current data and model dependencies, a joint analysis of the energy spectrum and anisotropy does not support the propagation origin model of the cosmic-ray knee at a 95\% confidence level. We hope that future LHAASO experiments will provide precise measurements of the energy spectra and anisotropies of various nuclei in the knee region, thereby offering a definitive test of the propagation model as the origin mechanism of the knee-region spectral structure. 

\end{abstract}


\section{Introduction}\label{sec1}

One of the most significant structures in the cosmic-ray (CR) energy spectrum is the so-called knee, a steepening feature around a few PeV, reminiscent of a human knee. While the overall spectrum from GeV to $\rm \sim50 ~EeV$ follows an approximate power-law behavior, precise measurements have revealed a series of fine structures. These include a spectral hardening near a rigidity of $\rm \sim 200 ~GV$ \citep{ATIC2007,ATIC2009,CREAM2010,PAMELA2011,CREAM2017,AMS2015prlA,AMS2015prlB,CALET2019,CALET2022}, softening of the proton and helium spectra at $\rm \sim14~TeV$ \citep{CREAM2017,NUCLEON2017,NUCLEON2018,DAMPE2019SciA} and $\rm \sim 34~TeV$ \citep{DAMPE2021prl} respectively, the knee at $\rm \sim3 ~PeV$ \citep{kulikov1958}, a second knee near 400 PeV \citep{Nagano1992,Bird1993prl,Lawrence1991,Glushkov2005,HiRes2008prl}, the ankle at $\rm \sim5 ~EeV$  \citep{HiRes2008prl,AUGER2008prl,TA2013apjl,AUGER2020prd}, and the GZK suppression above several tens of EeV \citep{GK1966prl,Z1966,HiRes2008prl}. Among these, the knee is particularly important, characterized by a change in spectral index from -2.7 to -3.1, and is widely considered to mark the transition between Galactic and extragalactic cosmic rays (EGCRs). Since its discovery in 1958, the knee has been a central focus in cosmic-ray physics, prompting extensive theoretical modeling efforts to explain its origin.

Current explanations for the knee fall into three main categories: source acceleration limits, propagation effects, and interaction scenarios. In the acceleration-limited models, the knee arises due to the maximum energy that cosmic-ray sources (e.g., supernova remnants) can provide \citep{Sveshnikova2003AA}, which scales linearly with charge number $Z$, implying a rigidity-dependent cutoff. Propagation-based models attribute the knee to changes in the diffusion properties of cosmic rays as they traverse the Galaxy, such as variations in the diffusion coefficient or escape time \citep{Berezinskii1990,Swordy1995ICRC,LAGUTIN2001,Ptuskin1993AA,Ogio2003,Roulet2004IJMPA,Candia2003JCAP,Candia2002JHEP033,Candia2002JHEP032}, also resulting in a rigidity-dependent feature. Alternatively, interaction-based models suggest that the knee is caused by interactions between cosmic rays and background particles (e.g., photons or neutrinos) \citep{KARAKULA1993,CANDIA2002,Dova2001}, or by new physics beyond the Standard Model in the Earth's atmosphere \citep{Nikolsky1999217,Nikolsky2000PAN}. These processes would result in a mass-number $A$-dependent knee structure.

Distinguishing among these scenarios requires determining whether the knee is primarily 
$Z-$, $R-$, or $A$-dependent \citep{Hehh2024}. If the knee depends on $A$, interaction-based origins are favored. However, a rigidity-dependent knee could result from either acceleration or propagation, making further discrimination challenging. Recent experiments have begun probing this dependence with improved precision.

In recent years, high-resolution measurements of the CR spectrum and anisotropy have offered new insights. The LHAASO experiment has precisely measured the all-particle spectrum in the $\rm 0.3–30 ~PeV$ range, clearly resolving the spectral break at the knee \citep{LHAASOspectrum2024prl}. Furthermore, LHAASO has reported high-precision measurements of the proton spectrum between 0.1 and 10 PeV, revealing the proton knee structure in detail \citep{2025arXiv250514447T}. IceTop has extended the proton spectrum above 10 PeV, while KASCADE and KASCADE-Grande have provided composition-sensitive measurements. KASCADE identified the knee as primarily due to a steepening of light components, though inconsistencies remain between different hadronic interaction models (e.g., QGSJet vs. SIBYLL)\citep{QGSJet1993PAN,SIBYLL1999ICRC,KASCADE2005}. KASCADE-Grande, using the QGSJET-II-02 model \citep{Ostapchenko2006143,Ostapchenko2006prd}, further identified a knee in the iron spectrum at $\rm \sim 80 ~PeV$ \citep{KASCADE2013}, suggesting the existence of a heavy-component knee.

If the knee arises from propagation effects, a sharp increase in anisotropy amplitude is expected near the knee energy. Recent anisotropy measurements by AS$\gamma$ \citep{2005ApJ...626L..29A,2017ApJ...836..153A,2015ICRC...34..355A}, KASCADE-Grande \citep{KASCADEAniso2015}, IceCube/IceTop \citep{2010ApJ...718L.194A,2012ApJ...746...33A,2013ApJ...765...55A,Aartsen2016,2025ApJ...981..182A} and the Pierre Auger Observatory (AUGER) \citep{2024ApJ...976...48A} have provided critical data. AUGER, in particular, has measured the anisotropy from tens of PeV to tens of EeV. Its results show a dip structure in anisotropy amplitude near $\rm \sim 1 ~EeV$ and a phase shift from Galactic to extragalactic directions \citep{AugerAniso2017Science,AugerAniso2018,AugerAniso2020}. Above $\rm \sim10 ~EeV$, cosmic rays appear to be of extragalactic origin.

Since both spectral features and anisotropies are shaped by the diffusion process, a combined analysis of these observables is essential for identifying the origin of the knee. If the knee arises from propagation effects, a correlation with anisotropy features is expected, whereas an acceleration-driven knee may not exhibit such behavior.

In this work, we investigate the propagation origin of the cosmic-ray knee by jointly analyzing spectral and anisotropy data. The paper is organized as follows: Section \ref{sec2} describes the methodology; Section \ref{sec3} presents the results; Section \ref{sec4} provides a summary and discussion. 

\section{Model and method}
\label{sec2} 
Cosmic rays originate from particle acceleration processes in extreme astrophysical environments, such as supernova remnants, rapidly rotating pulsars, intense stellar winds from massive stars, and energetic microquasar jets~\citep{2024arXiv241008988L,2024arXiv241213889Y,2025PhRvD.112l3015Z}. These high-energy astrophysical objects accelerate charged particles to nearly the speed of light through mechanisms like shock acceleration and magnetic reconnection, before ejecting them into the vast interstellar medium. These particles then undergo diffusion through the Galactic magnetic field for millions of years, continuously interacting with interstellar matter and modifying their spectral properties. The propagation of cosmic rays is precisely described by the following equation \citep{Strong2007}:
\begin{equation}
\label{Eq2.1}
\begin{aligned}
\frac{\partial \psi(\vec{r}, p, t)}{\partial t}= & q(\vec{r}, p, t)+\vec{\nabla} \cdot\left(D \vec{\nabla} \psi-\vec{V} \psi\right) \\
& +\frac{\partial}{\partial p} p^2 D_{p} \frac{\partial}{\partial p} \frac{1}{p^2} \psi-\frac{\partial}{\partial p}\left[\dot{p} \psi-\frac{p}{3}(\vec{\nabla} \cdot \vec{V}) \psi\right] \\
& -\frac{1}{\tau_f} \psi-\frac{1}{\tau_r} \psi.
\end{aligned}
\end{equation}

Here, $\psi(\vec{r}, p, t)$ represents the density of cosmic rays with momentum $p$ at position $\vec{r}$, and $q(\vec{r}, p, t)$ denotes the source term. D is the spatial diffusion coefficient, $\vec{V}$ is the convection velocity, $D_{p}$ is the diffusion coefficient in momentum space, $\tau_{f}$ is the timescale for fragmentation, and $\tau_{r}$ is the timescale for radioactive decay. Therefore, cosmic rays reaching the Earth primarily undergo two key stages: an initial acceleration phase in the source region, followed by long-term diffusion throughout the Galactic scale.

\subsection{Distribution and Injection Spectrum of Sources}
The source term $q(\vec{r}, p, t)$ describes the spatial distribution and injection power spectrum of cosmic ray sources, characterizing both the source locations and the momentum spectrum of injected particles. As the primary sources of Galactic cosmic rays, supernova remnants (SNRs) are widely used to represent the source distribution pattern within the Galaxy. In cylindrical coordinates, this distribution can be expressed as \citep{Case1996AAPS}:
\begin{equation}
\label{Eq2.2}
f(r, z) \propto\left(r / r_{\odot}\right)^{1.69} \exp \left[-3.33\left(r-r_{\odot}\right) / r_{\odot}\right] \exp \left(-|z| / z_s\right), 
\end{equation}
where the radial distance of the Sun is $r_{\odot}=\rm 8.5~kpc$, and the vertical scale height is $z_{s}=\rm 0.2~kpc$.

After undergoing acceleration in source regions (such as SNR shocks), cosmic ray particles are injected into the interstellar medium. A key feature of this acceleration mechanism is the resulting power-law energy spectrum, where the particle flux exhibits a dependence on energy described as \citep{horandel_knee_2003,Guo2016apj}
\begin{equation}
\label{Eq2.3}
    q\left(\mathcal{R}\right)=q_{0}\times \left\{\begin{array}{ll}
         \left(\frac{\mathcal{R}}{\mathcal{R}_{br}}\right)^{-\nu_{1}}, & \mathcal{R}< \mathcal{R}_{br}  \\
         \left(\frac{\mathcal{R}}{\mathcal{R}_{br}}\right)^{-\nu_{2}} , & \mathcal{R}\ge\mathcal{R}_{br}
    \end{array}\right.,
\end{equation}
where $\mathcal{R}=p/Ze$ is the particle rigidity, $q_0$ is the normalization factor for all nuclei, $\mathcal{R}_{br}$ is the break rigidity, and $\nu_1$, $\nu_2$ are the spectral indices below and above the break, respectively.

It is worth noting that in acceleration-limit models explaining the origin of the cosmic ray knee, an exponential suppression factor $e^{-\mathcal{R}/\mathcal{R}_{c}}$ is often introduced to account for the observed knee feature in the spectrum. Here, $\mathcal{R}_c$ represents the cutoff rigidity, physically interpreted as the upper limit of acceleration capability of Galactic sources. This explanation corresponds to the so-called Z-dependent acceleration origin of the cosmic ray knee. However, since this study focuses on the propagation origin of the knee, this factor is omitted from our baseline model.

To account for the observed hardening in the cosmic ray spectrum near 200 GeV and the softening around 14 TeV, we introduce a nearby source component into the model. This additional source significantly improves the model’s ability to reproduce spectral features below 100 TeV. Similar to previous work, the location of the source is set at  $l=161\degree$, $b=9\degree$\citep{2019JCAP...10..010L}. The spectrum of the source follows a broken power-law form, similar to Eq.\ref{Eq2.3}, and is given by:

\begin{equation}
\label{Eq2.4}
    q_{\rm local}\left( \mathcal{R}\right)=q_{\rm 0,local}\cdot 
         \left(\frac{\mathcal{R}}{\mathcal{R}_{br}}\right)^{-\beta}  \cdot  e^{-\mathcal{R}/\mathcal{R}_{\rm c,local}}, 
\end{equation}
where the exponential term $e^{-\mathcal{R}/\mathcal{R}_{\rm c,local}}$ represents the acceleration limit of the nearby source with $\rm \mathcal{R}_{c,local}=20~TeV$.

After acceleration at source regions such as SNR shocks, cosmic rays are injected into the interstellar medium and begin diffusing throughout the Galaxy. During this propagation phase, a portion of the cosmic rays ultimately reaches Earth and is detected. This diffusion process is primarily governed by the Galactic diffusion coefficient, which characterizes the transport properties of cosmic rays within the interstellar magnetic field.

\subsection{Propagation Coefficient}
The diffusive behavior of cosmic rays can be quantitatively described by the diffusion coefficient $\rm D(E)$, which is a function of the particle energy $\rm E$ and characterizes the transport properties of cosmic rays in the interstellar magnetic field. According to diffusion theory, $\rm D(E)$ typically follows a power-law dependence, with 
$\delta$ representing the diffusion index. The value of $\delta$ is closely related to the turbulence properties of the interstellar medium. The specific form is given by:

\begin{equation}
    \label{Eq2.5}
    D=\left\{\begin{array}{ll}
         D_0\beta^{\varepsilon}\left(\frac{\mathcal{R}}{\mathcal{R}_{0}}\right)^{\delta}, & \mathcal{R}< \mathcal{R}_{knee}  \\
         D_0\beta^{\varepsilon}\left(\frac{\mathcal{R}}{\mathcal{R}_{0}}\right)^{\delta}\left(\frac{\mathcal{R}}{\mathcal{R}_{knee}}\right)^{\delta_{knee}}, & \mathcal{R}\ge\mathcal{R}_{knee}
    \end{array}\right.,
\end{equation}
where $D_{0}=4.87\times10^{28}\rm~cm^{2}~s^{-1}$ is the normalized diffusion coefficient at the reference rigidity $\mathcal{R}_{0}=4\rm~GeV$, $\beta=v/c\sim1$ is the ratio of the cosmic-ray speed to light speed, and $\varepsilon=0.05$. 

In particular, the term $\left(\frac{\mathcal{R}}{\mathcal{R}_{knee}}\right)^{\delta_{knee}}$ accounts for the sudden change in the diffusion coefficient across the knee region. 
$\mathcal{R}_{knee}$ is the critical rigidity corresponding to the cosmic ray knee, and 
$\delta_{knee}$ describes the change in the spectral index of the diffusion coefficient before and after the knee. In the propagation-origin model, the source injection spectrum does not include a cutoff in Eq.~\ref{Eq2.3}, i.e., the cutoff rigidity $\mathcal{R}_{c}\to \infty$. From the perspectives of both the acceleration-origin and propagation-origin models, their predictions for the cosmic ray spectrum in the knee region are indistinguishable based solely on spectral observations. Therefore, the spectral data alone are insufficient to differentiate between the two models. However, their predictions for anisotropy are distinct, making cosmic ray anisotropy measurements a crucial tool for distinguishing and testing these models in our study.

\subsection{Anisotropy of Cosmic Rays}
The key to this study lies in accurately determining the two parameters, $\rm \mathcal{R}_{knee}$ and $\rm \delta_{knee}$, in the propagation-origin model by utilizing experimental observations of both the cosmic ray spectra in the knee region and cosmic ray anisotropy. To achieve this, we must first ensure that the other parameters in Eq.~\ref{Eq2.5} are accurately constrained, which requires adopting a low-energy propagation model that is well-aligned with experimental measurements below the knee. For this purpose, we adopt a well-developed class of spatially dependent propagation models, namely the SDP (Spatially Dependent Propagation) model.

In the SDP model, to explain the spectral hardening at a rigidity of about 200 GeV and the softening near an energy of 14 TeV, a spatially dependent diffusion coefficient is introduced. That is, both $\rm D_0$ and $\delta$ in Eq.\ref{Eq2.5} are assumed to vary with the spatial source distribution described by Eq.\ref{Eq2.2}. Additionally, a nearby source located in the anti-Galactic-center direction is introduced. This not only explains the dip structure observed in the amplitude of the anisotropy from approximately 
200 GeV to the knee region but also accounts for the phase flip in the anisotropy—observed around 100 GeV—from the anti-Galactic-center direction to the Galactic center direction. For further details, see \citep{Guo2018prd,Liu2019JCAP,Qiao2019JCAP}.

Based on this framework, we can extract the parameter ranges of $\rm \mathcal{R}_{knee}$ and $\rm \delta_{knee}$ from experimental measurements of the cosmic ray spectrum and anisotropy above the knee region. By comparing these two sets of results, we can further test the validity of the model.

However, at energies beyond the knee, the observed anisotropy of ultra-high-energy cosmic rays inevitably includes contributions from extragalactic sources. This makes it difficult for our model—which focuses on Galactic propagation—to accurately interpret the experimental anisotropy data. Therefore, before comparing with anisotropy measurements, we need to extract the Galactic component of the anisotropy from the ultra-high-energy cosmic ray data.

We provide a detailed method for calculating the amplitude of the Galactic cosmic ray anisotropy in \ref{appendixA}. Based on this method and the data from the AUGER experiment, we present the anisotropy of ultra-high-energy Galactic cosmic rays in \ref{appendixB}.

\begin{figure}[t]
\centering
\includegraphics[width=0.95\linewidth]{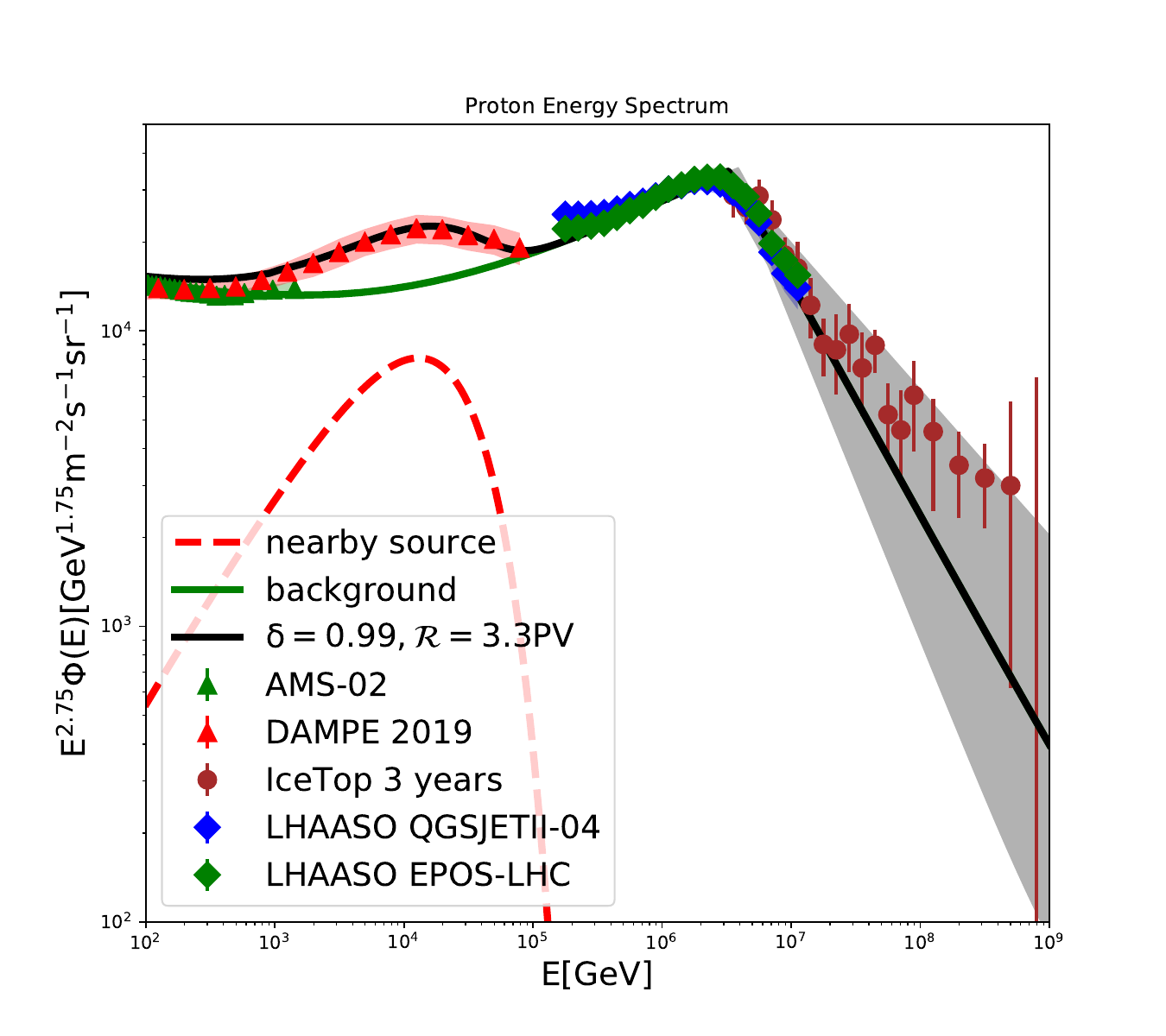}
\caption{The calculated proton spectrum compared with observed data. The data points are taken from the AMS-02 \citep{2015PhRvL.114q1103A}, DAMPE-2019 \citep{2019SciA....5.3793A}, IceTop \citep{2019PhRvD.100h2002A}, and LHAASO \citep{2025arXiv250514447T}. The gray shaded area represents the $1~\sigma$ uncertainty band of the energy spectrum calculated based on the $1~\sigma$ errors of $\rm \mathcal{R}_{knee}$ and $\rm \delta_{knee}$.}
\label{fig1}
\end{figure}

\begin{figure}[t]
\centering
\includegraphics[width=0.95\linewidth]{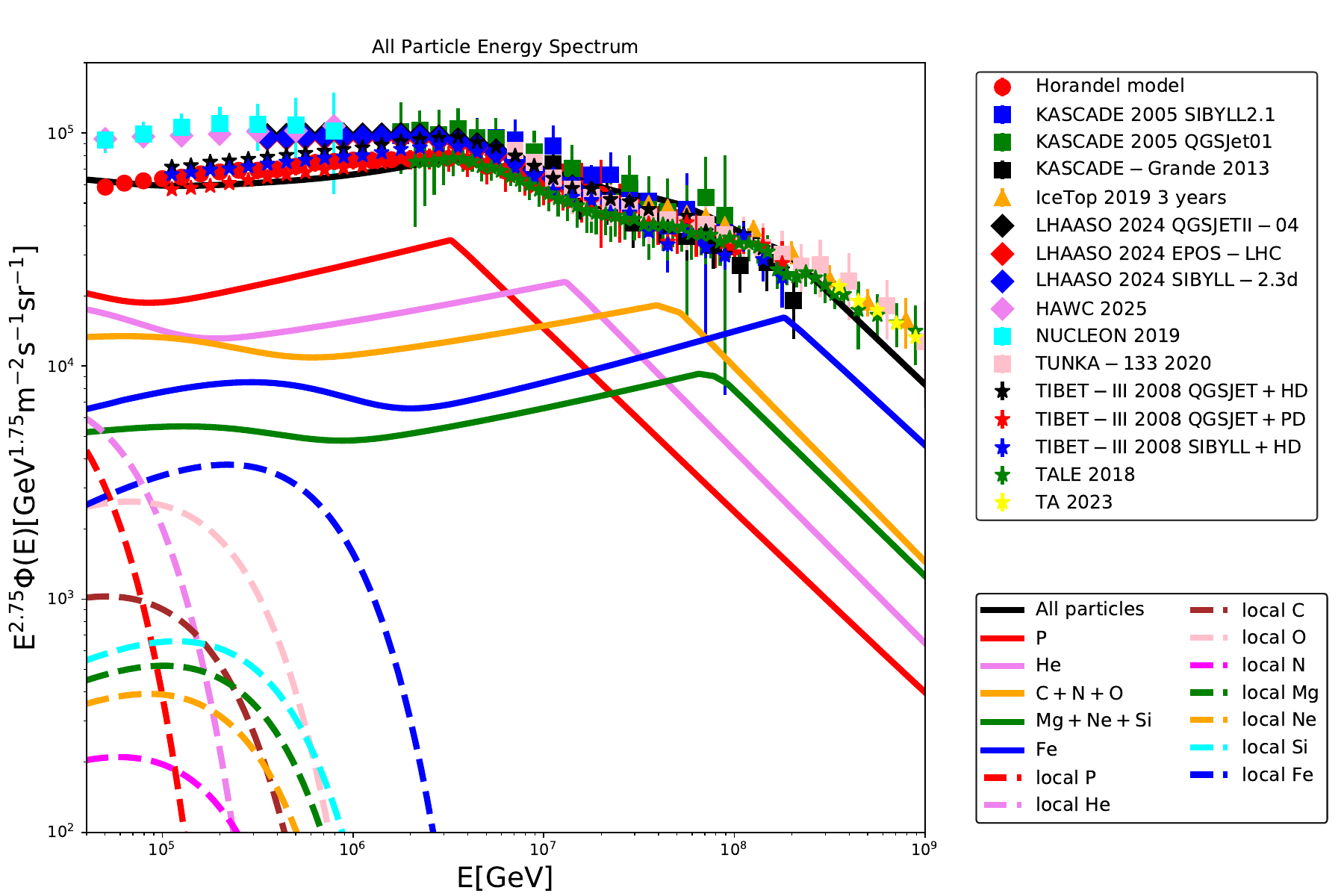}
\caption{Sum flux of individual species, compared with all-particle spectrum data from experiments. All-particle spectra are from Tibet-III \citep{2008ApJ...678.1165A}, TALE \citep{2018ApJ...865...74A}, TA \citep{2023APh...15102864A}, NUCLEON \citep{2019AdSpR..64.2546G}, KASCADE \citep{2024JCAP...05..125K}, HAWC \citep{2017PhRvD..96l2001A}, KASCADE-Grande \citep{2013APh....47...54A}, TUNKA-133 \citep{2020APh...11702406B} and LHAASO \citep{LHAASOspectrum2024prl}.}
\label{fig2}
\end{figure}

\section{Results}\label{sec3}
We incorporate the diffusion coefficient expression Eq.~\ref{Eq2.5} constructed in our method into DRAGON~\citep{2017JCAP...02..015E} (Diffusion of cosmic RAys in galaxy modelizatiON, a numerical code used to resolve the cosmic ray propagation equation) to calculate cosmic-ray energy spectra and anisotropy. The SDP model is simultaneously employed to reproduce low-energy experimental observations, thereby constraining model parameters other than $\rm \delta_{knee}$ (the difference in spectral indices before/after the knee) and $\rm \mathcal{R}_{knee}$ (the knee position).
The specific values of $\rm \delta_{knee}$ and $\rm \mathcal{R}_{knee}$ have been precisely determined by recent LHAASO observations as $\rm \delta_{knee}=0.99\pm 0.29$ and $\rm \mathcal{R}_{knee} =3.3\pm 0.6~PV$ \citep{2025arXiv250514447T}. Assuming the knee structure originates from propagation effects, we replicate the observed knee-region spectral structure in the LHAASO proton spectrum and calculate the corresponding anisotropy to evaluate the confidence level of this propagation origin hypothesis.

\begin{figure}[t]
\centering
\includegraphics[width=0.95\linewidth]{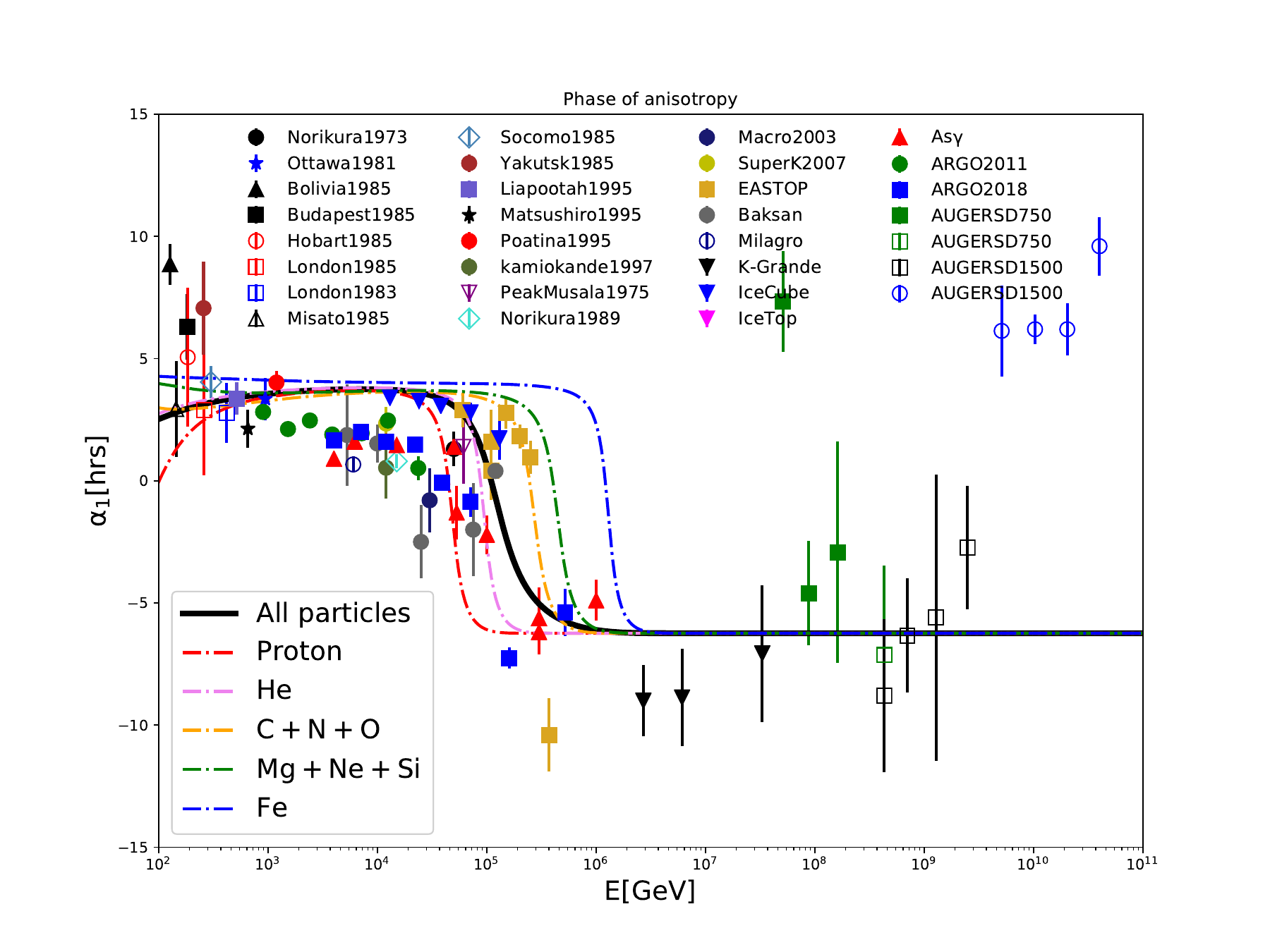}
\caption{The energy dependence of the phases of the dipole anisotropies when adding all of the major elements together. The data points are taken from Norikura \citep{1973ICRC....2.1058S}, Ottawa \citep{1981ICRC...10..246B}, Bolivia \citep{1985P&SS...33.1069S}, Budapest \citep{1985P&SS...33.1069S}, Hobart \citep{1985P&SS...33.1069S}, London \citep{1985P&SS...33.1069S}, Misato \citep{1985P&SS...33.1069S}, Socomo \citep{1985P&SS...33.1069S}, Yakutsk \citep{1985P&SS...33.1069S}, Liapootah \citep{1995ICRC....4..639M}, Matsushiro \citep{1995ICRC....4..648M}, Poatina \citep{1995ICRC....4..635F}, kamiokande1 \citep{1997PhRvD..56...23M}, PeakMusala \citep{1975ICRC....2..586G}, Norikura \citep{1989NCimC..12..695N}, Macro \citep{2003PhRvD..67d2002A}, SuperK \citep{2007PhRvD..75f2003G}, EAS-TOP \citep{1995ICRC....2..800A,1996ApJ...470..501A,2009ApJ...692L.130A}, Baksan \citep{1987ICRC....2...22A}, Milagro \citep{2009ApJ...698.2121A}, K-Grande \citep{KASCADEAniso2015}, IceCube \citep{2010ApJ...718L.194A,2012ApJ...746...33A,2025ApJ...981..182A}, IceTop \citep{2013ApJ...765...55A}, AS-$\gamma$ \citep{2005ApJ...626L..29A,2017ApJ...836..153A,2015ICRC...34..355A}, ARGO \citep{2018ApJ...861...93B}, AUGER \citep{2024ApJ...976...48A}.}
\label{fig3}
\end{figure}

\begin{figure}[t]
\centering
\includegraphics[width=0.95\linewidth]{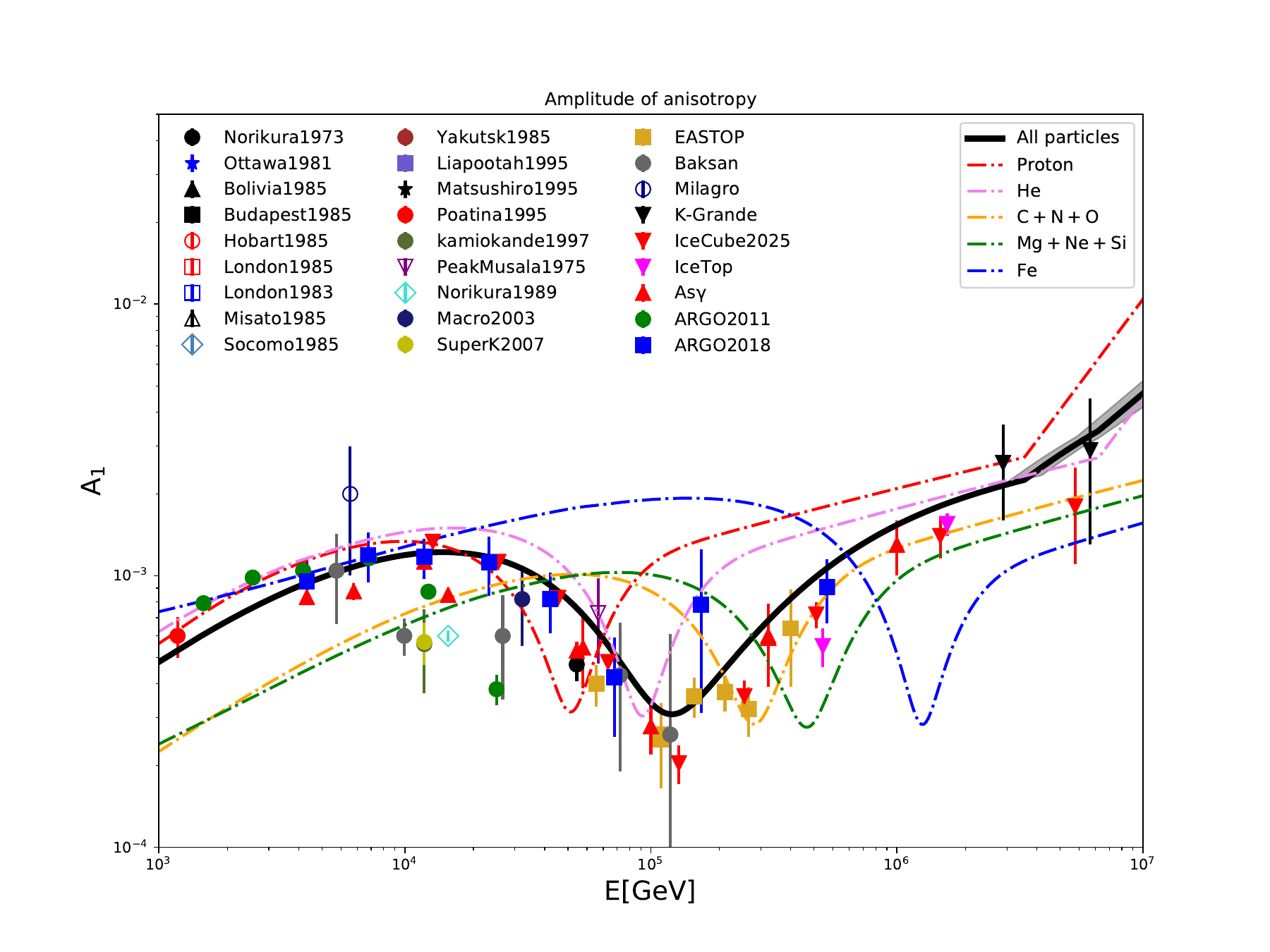}
\includegraphics[width=0.95\linewidth]{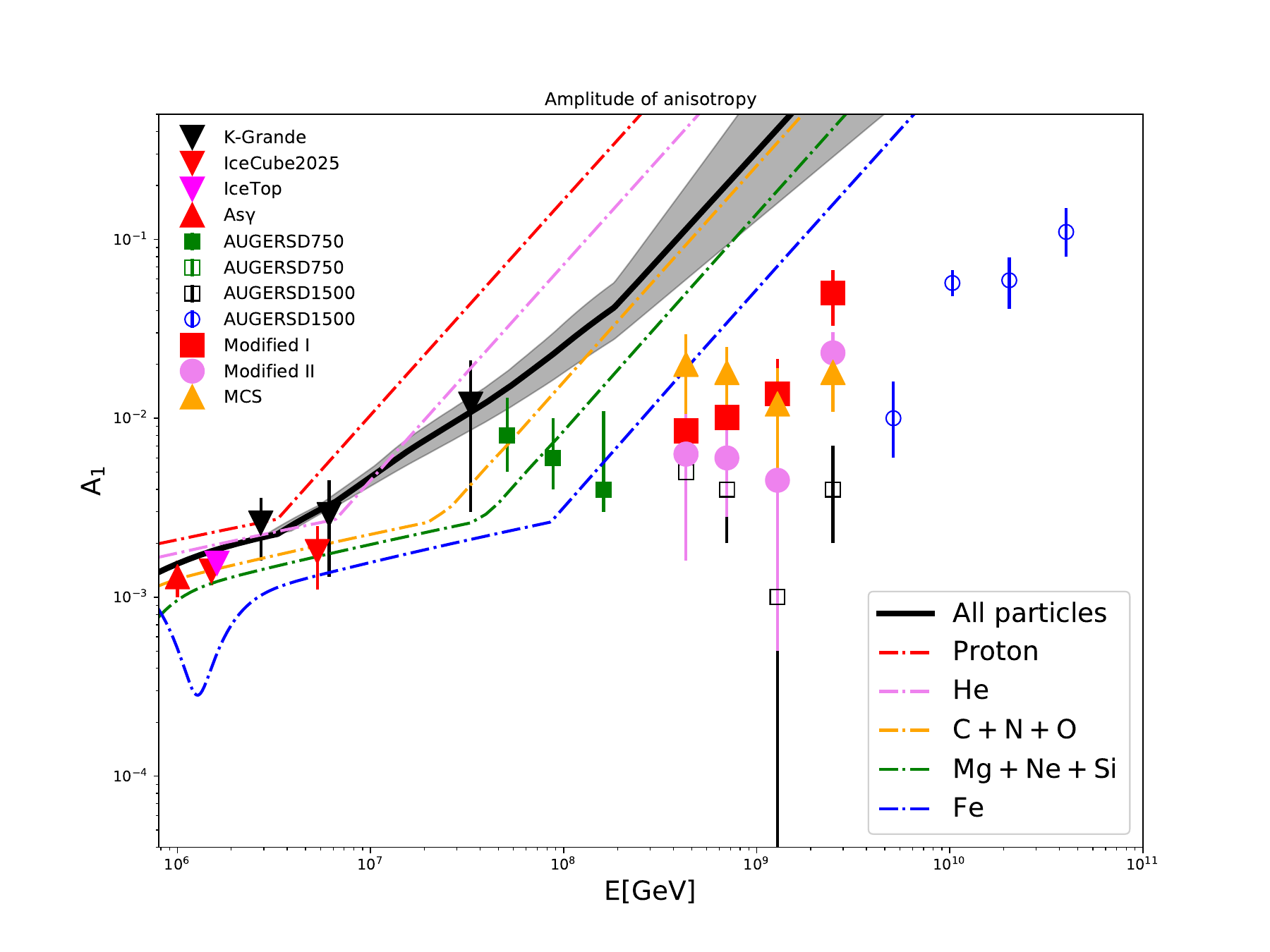}
\caption{The same as Fig.~\ref{fig3} but for the amplitudes. The gray shaded area represents the $1~\sigma$ uncertainty band of the amplitude of anisotropy calculated based on the $1~\sigma$ errors of $\rm \mathcal{R}_{knee}$ and $\rm \delta_{knee}$, which have been precisely determined by LHAASO proton energy spectrum \citep{2025arXiv250514447T}. The amplitude of anisotropy with energies below 10 PeV is shown in the upper panel, while the amplitude of anisotropy with energies above 1 PeV is plotted in the lower panel. In the lower panel, $\rm Modified-I$: Contamination correction applied using only the last three high-energy (blue) data points to adjust low-energy (black) data; $\rm Modified-II$: Comprehensive correction using all high-energy (blue) data points; MCS: the most conservative scenario, giving upper limits of GCR amplitude constrained by upper limits of EGCR amplitude, see Table \ref{tab3} in \ref{appendixC}.}
\label{fig4}
\end{figure}

\subsection{CR Spectra}
Figure \ref{fig1} displays the fitting results for the cosmic-ray proton spectrum. The ``bump" structure observed in the $\rm \sim 200 ~GeV$ to $\rm 14 ~TeV$ energy range is attributed to a local source (Eq.\ref{Eq2.4}), while the spectral break at the knee region is successfully reproduced by introducing a sudden break in the diffusion coefficient at the knee energy (Eq.\ref{Eq2.5}).

Using the same methodology, we calculate heavier cosmic-ray nuclei and combine all components to obtain the all-particle spectrum shown in Figure \ref{fig2}. The first and second knees are dominated by the proton knee and iron nucleus knee components respectively, with the superposition of nuclear contributions collectively shaping the spectrum between these two knees.

By adjusting parameters from a propagation perspective, we effectively reproduce both the cosmic-ray proton spectrum and the all-particle spectrum. Under this framework, the consistency between the model-predicted cosmic-ray anisotropy values and observational data serves as a critical test for verifying whether the knee structure originates from propagation processes.

\subsection{CR Anisotropy}
Figure \ref{fig3} compares the model-predicted phase of cosmic-ray anisotropy with observational data. Except for the AUGERSD1500 dataset, the model provides reasonably accurate predictions for observations at energies below AUGERSD1500. Notably, in AUGERSD1500 data:
Particles above $\rm \sim5 ~EeV$ exhibit exclusive alignment toward the anti-Galactic center direction, suggesting a predominantly extragalactic origin.
Data below $\rm \sim5 ~EeV$ are dominated by Galactic cosmic rays (GCRs) but contaminated by an extragalactic component.
Since our model describes exclusively Galactic-origin cosmic rays, we apply an extragalactic contamination correction to the AUGERSD1500 anisotropy amplitude data (detailed in \ref{appendixB}) for meaningful theory-experiment comparison. The corrected results (Figure \ref{fig4}) show:
$\rm Modified-I$: Contamination correction applied using only the last three high-energy (blue) data points to adjust low-energy (black) data;
$\rm Modified-II$: Comprehensive correction using all high-energy (blue) data points; the most conservative scenario (MCS): giving upper limits of GCR amplitude constrained by upper limits of EGCR amplitude, see Table \ref{tab3} in \ref{appendixC} for details.

Figure \ref{fig4} directly contrasts observed anisotropy amplitudes with model predictions:
Upper panel (energy $\rm < 10 ~PeV$): Strong agreement between model and observations.
Lower panel: Significant deviation of model predictions from data.
Using hypothesis testing (methodology in \ref{appendixC}), we quantify the agreement between predicted and observed anisotropy amplitudes from 2.7 PeV to 2.48 EeV. The model predictions reject the null hypothesis (that the knee structure arises from propagation processes) at 95\% confidence level (CL).

It is worth noting that the confidence level we calculate above is a rather conservative estimate. The corrections we have made, namely $\rm Modified-I$ and $\rm Modified-II$, are intended to extract the anisotropy amplitude of Galactic cosmic rays (GCRs) from experimental observations contaminated by extragalactic cosmic rays (EGCRs), in order to compare them with our theoretical calculations. The anisotropy observations near the EeV energies (Figure 4, black square data points) are completely dominated by GCRs, as their phase points toward the Galactic center (Figure 3, black square data points). Furthermore, even in the most conservative scenario (MCS), which is independent of the composition model of ultra-high-energy cosmic rays, the calculated confidence level remains 95\% (see \ref{appendixC} for details). Consequently, our results rule out a propagation origin for the cosmic-ray knee structure at 95\% CL based on anisotropy observations.

Future higher-precision measurements of ultra-high-energy cosmic-ray anisotropy and spectra will provide critical tests of this conclusion.

\section{Conclusion}\label{sec4}
\label{sec:Conclusion and Discussion}
The origin of the cosmic-ray knee structure remains unresolved. Proposed explanations—including acceleration origins, propagation origins, and interaction origins—have yet to definitively identify or exclude which theory aligns closest with physical reality.

Recent years have brought more precise experimental measurements of energy spectra and anisotropy at and above the knee region, offering new hope for determining the knee's origin.

Building on the latest observational data for cosmic-ray spectra and anisotropy, this work jointly investigates and tests the propagation origin hypothesis for the knee structure. Our approach involves:
(1) Reconstructing the diffusion coefficient to reproduce the proton knee spectrum observed by LHAASO and the all-particle knee structure from multi-experiment data;
(2) Calculating cosmic-ray anisotropy and phase distributions.

Through hypothesis testing, we find that propagation-origin models for the knee structure are ruled out at the 95\% confidence level by anisotropy measurements.

It should be noted that our analysis relies on a key assumption: the extragalactic anisotropy component observed by AUGER in the southern hemisphere would properly represent its counterpart in the northern hemisphere, where LHAASO is located. However, there could be several discrepancies between the northern and southern hemispheres. Consequently, if one had used the southern hemisphere anisotropies for eliminating the EG component in the northern hemisphere, the subtraction may not be close to correct. We have to acknowledge that the 95\% CL conclusion might need to change if the extrapolation of the southern hemisphere EG anisotropy results to the northern-hemisphere anisotropy background is not fully valid.

Given that current ultra-high-energy anisotropy data still carry substantial uncertainties, we anticipate that future precision measurements will provide more robust validation for these conclusions.

\section*{Acknowledgements}
This work is supported in China supported by National Key R\&D program of China under the grant 2024YFA1611402 and the National Natural Science Foundation of China (12333006,12275279).

\begin{appendices}  
\renewcommand{\thesection}{Appendix \Alph{section}}
\section{Calculation Methods for UHE GCR Anisotropy}\label{appendixA}
Experimental observations of ultra-high-energy cosmic ray anisotropy inevitably contain contributions from extragalactic cosmic rays. Since our model exclusively describes Galactic cosmic rays (EGCRs), we must extract the Galactic component from the observed anisotropy. The cosmic ray anisotropy is conventionally defined as:

\begin{equation}
\label{EqA.1}
\vec{\delta}=\frac{3D}{v}\frac{\nabla\psi}{\psi},\tag{A.1}
\end{equation}
where $D$ is the diffusion coefficient, $v$ the particle velocity, and $\psi$ the cosmic ray flux.

Given that observations measure a mixture of Galactic ($\psi_G$) and extragalactic ($\psi_E$) components, we establish: 
\begin{equation}
\label{EqA.2}
\left\{\begin{array}{ll}\psi_{G}=\psi_{tot}-\psi_{E}\\
    \vec{\delta}_{G}\psi_{G}=\vec{\delta}_{tot}\psi_{tot}-\vec{\delta}_{E}\psi_{E}\end{array}\right.,\tag{A.2}
\end{equation}
where
$\psi_{tot}$ is observed total cosmic ray flux,
$\psi_{E}$ is observed extragalactic component flux,
$\vec{\delta}_{tot}$ is observed anisotropy vector,
$\vec{\delta}_{E}$ is extragalactic anisotropy vector, and
$\vec{\delta}_{G}$ is Galactic anisotropy vector.

$\vec{\delta}_{G}$ is derived as:
\begin{equation}
\label{EqA.3}
\vec{\delta}_{G}=\frac{\vec{\delta}_{tot}-\eta\vec{\delta}_{E}}{1-\eta},\tag{A.3}
\end{equation}
where:
\begin{equation}
\label{EqA.4}
\eta=\psi_{E}/\psi_{tot}.\tag{A.4}
\end{equation}
$\eta$ represents the fractional contribution of extragalactic cosmic rays to the total observed flux.

Notably, the phase of extragalactic anisotropy is nearly opposite to that of Galactic anisotropy. From Eq. \ref{EqA.3}, we obtain the Galactic anisotropy amplitude:
\begin{equation}
\label{EqA.5}
\delta_{G}=\frac{\delta_{tot}+\eta\delta_{E}}{1-\eta}.\tag{A.5}
\end{equation}

\begin{figure}[t]
\centering
\includegraphics[width=0.95\linewidth]{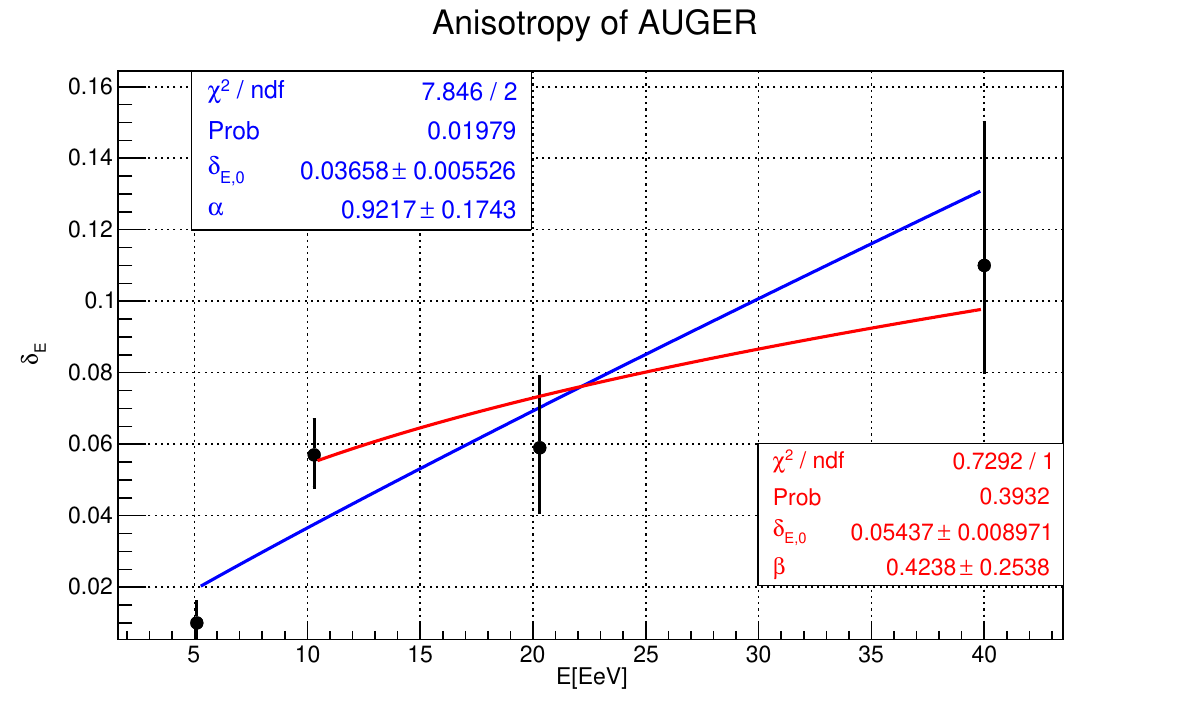}
\caption{Fitting of extragalactic cosmic ray anisotropy observed by the AUGER experiment\citep{2024ApJ...976...48A}. The red line represents the fitting result of the three highest energy data points, and the blue line represents the fitting result of all four data points.}
\label{fig5}
\end{figure}

\begin{figure}[t]
\centering
\includegraphics[width=0.95\linewidth]{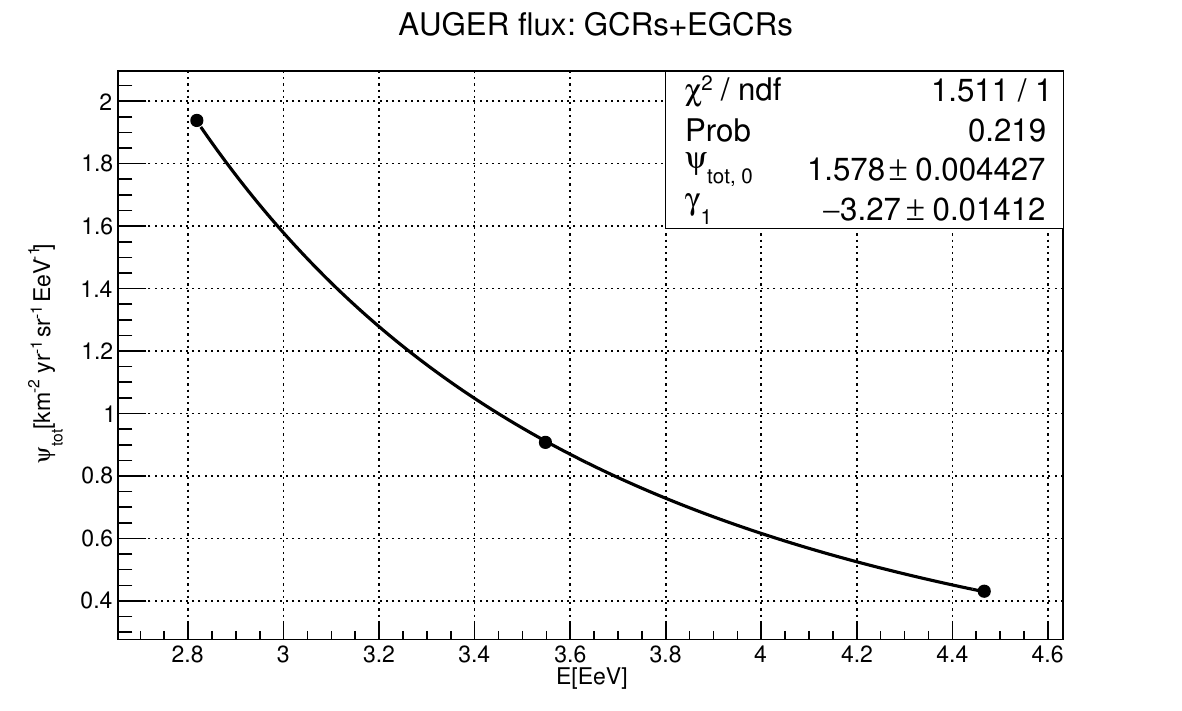}
\includegraphics[width=0.95\linewidth]{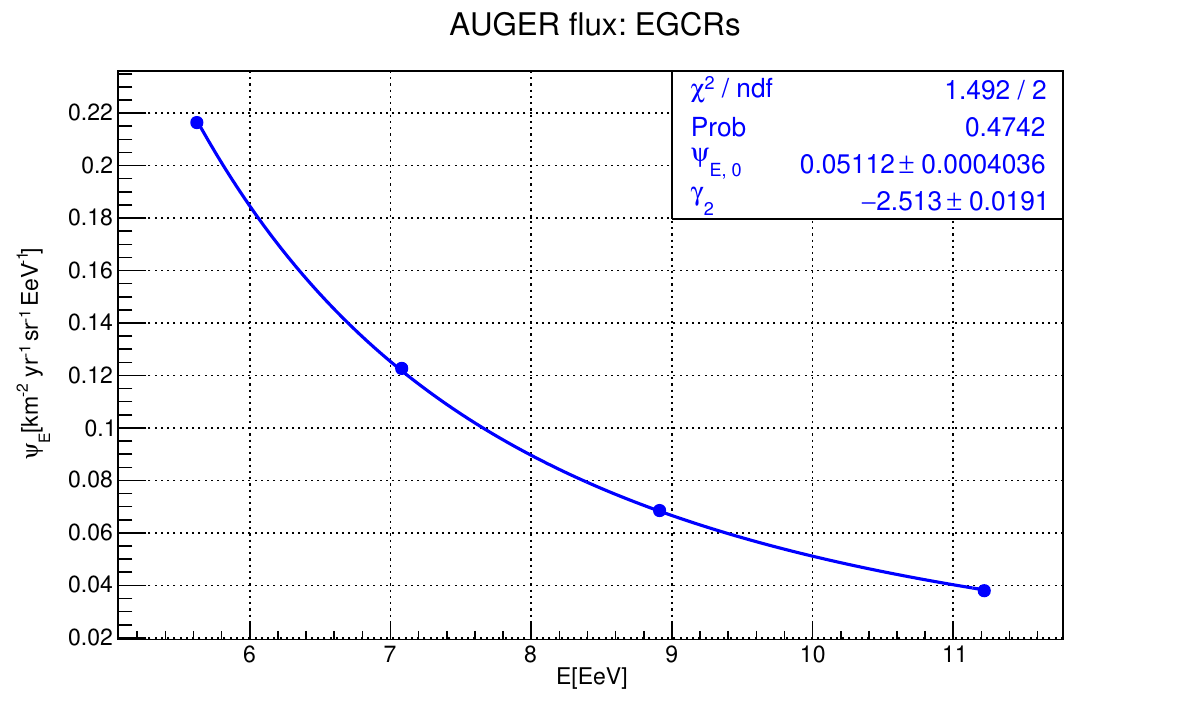}
\caption{Fitting near the ``ankle" structure of the cosmic-ray energy spectrum measured by AUGER \citep{AUGERspectrum}.}
\label{fig6}
\end{figure}

\begin{figure}[t]
\centering
\includegraphics[width=0.95\linewidth]{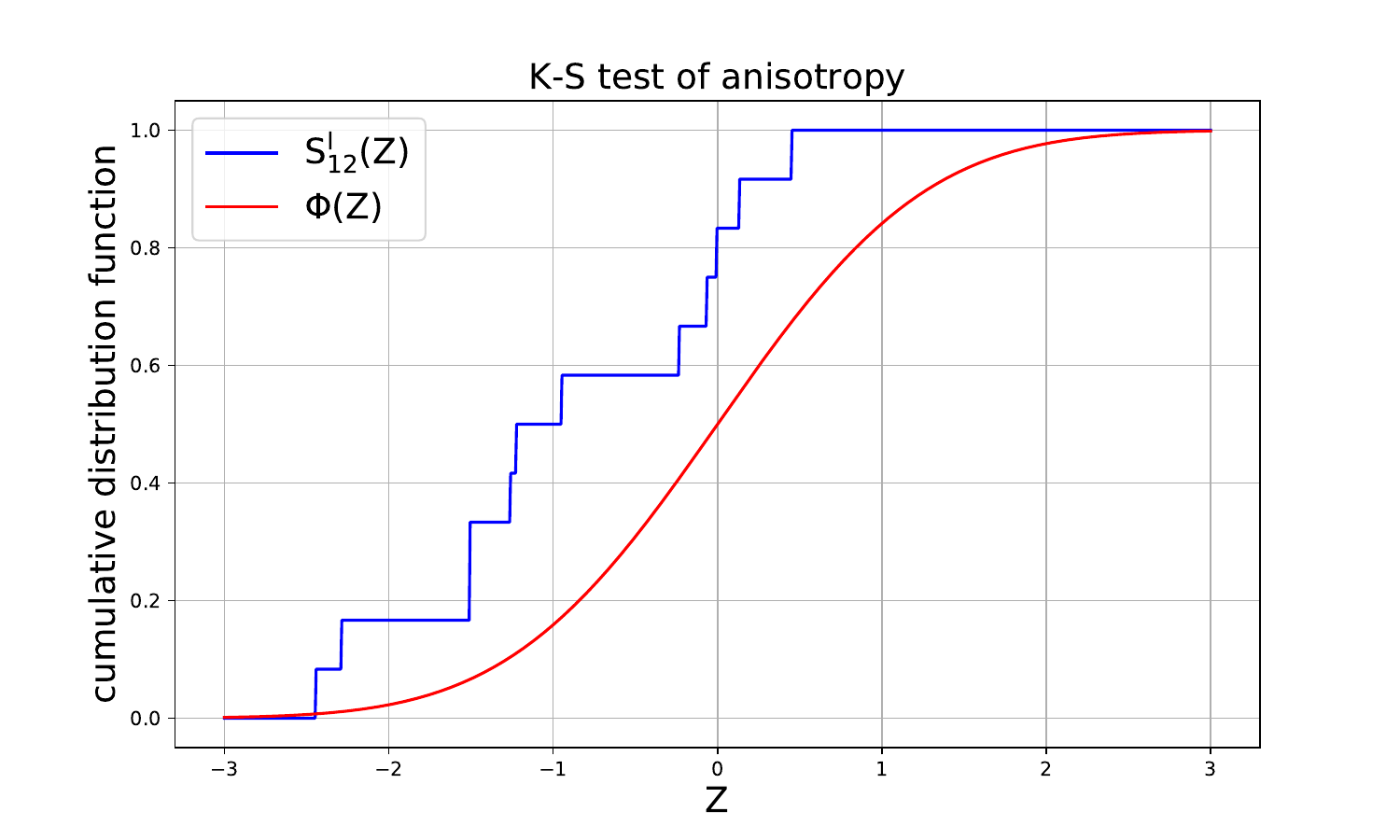}
\includegraphics[width=0.95\linewidth]{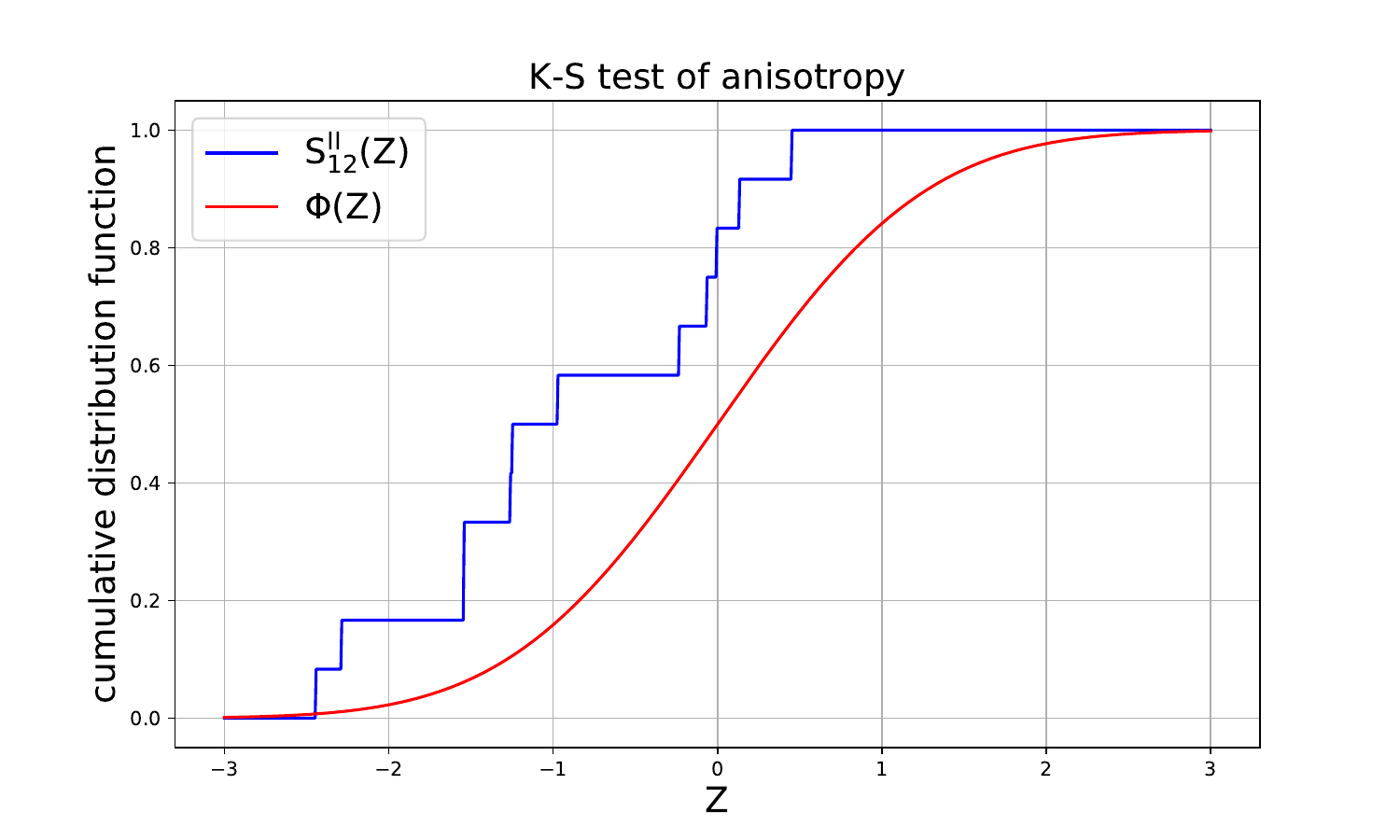}
\includegraphics[width=0.95\linewidth]{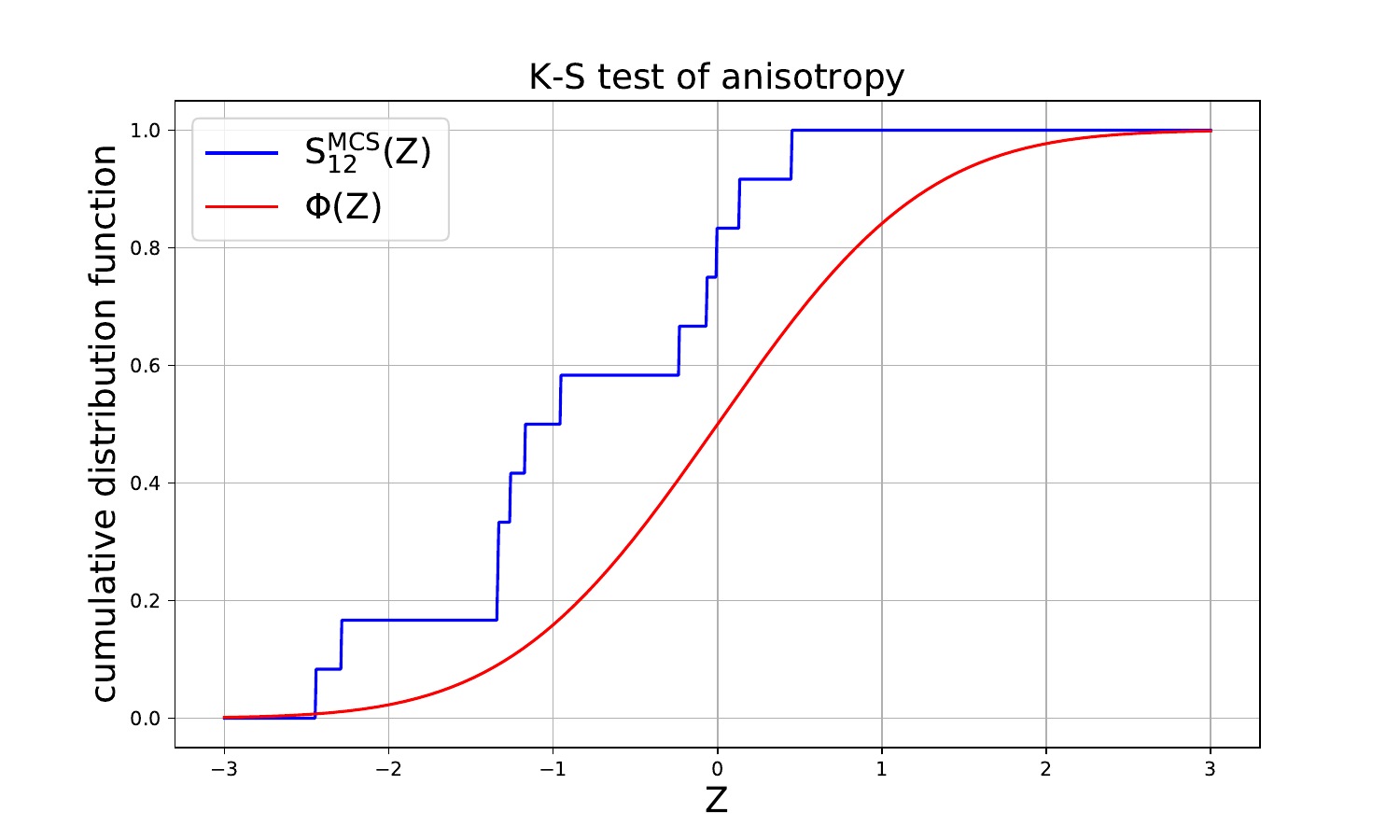}
\caption{Top panel: Kolmogorov test results using Modified I for AUGER data, middle panel: Kolmogorov test results using Modified II for AUGER data, bottom panel: Kolmogorov test results for the most conservative scenario (MCS).}
\label{fig7}
\end{figure}

\section{Anisotropy amplitude of GCRs from the AUGER dip structure}\label{appendixB}
Based on the anisotropy measurements and the all-particle energy spectrum from the AUGER, we can calculate the amplitude of GCR anisotropy at the AUGER dip structure using Eq \ref{EqA.5}. The phase of the anisotropy measured by AUGER shifts from the Galactic Center direction to the anti-Galactic Center direction between $\rm 10^9 ~GeV$ and $\rm 10^{10} ~GeV$, suggesting a transition from minor contamination to dominance by extragalactic cosmic rays. The opposing phases of extragalactic and galactic cosmic rays lead to a partial cancellation of their anisotropy amplitudes, thereby explaining the dip structure observed by AUGER around EeV energies.

Turning to the energy spectrum measured by AUGER, we note that the ``ankle" feature occurs at energies similar to the dip in anisotropy amplitude, indicating a potential connection. Combining these observational features of anisotropy and spectrum, we infer that the flux above the ankle is dominated by EGCRs, while the region below the ankle represents a mixture of galactic and extragalactic components. This framework allows all physical quantities required for Eq \ref{EqA.5} to be determined.

First, we fit the energy dependence of the extragalactic anisotropy amplitude measured by AUGER and extrapolate it to the dip structure. The phase confirms that the four highest-energy data points are of extragalactic origin, while their amplitudes suggest that the lowest-energy point among these four may still retain some galactic contamination. To conservatively address this, we performed separate fits using both the four highest-energy points and the three highest-energy points, employing the function $\delta_{E}=\delta_{E,0}\left(\frac{\mathrm{E}}{10 \mathrm{EeV}}\right)^\alpha$. The fitting results are shown in figure \ref{fig5}, and the extrapolating results are listed in the table \ref{tab1} and \ref{tab2}. Here, the error of extrapolating data are obtained by $\rm \sigma_{\delta_{E}}=\delta_{E}\sqrt{\left(\frac{\sigma_{\delta_{E,0}}}{\delta_{E,0}}\right)^2+\left(\sigma_\alpha \ln \frac{E}{10~ \mathrm{EeV}}\right)^2}$.

Then, we separately fitted the two segments of the energy spectrum near the ``ankle" structure measured by AUGER, extrapolating them to the dip structure to obtain the total cosmic-ray flux and the extragalactic cosmic-ray flux, respectively. The fitting results for these two spectral segments are shown in Figure \ref{fig6}. The extrapolating results ($\rm \psi_{tot}$ and $\rm \psi_{E}$) are presented in table \ref{tab2}. The fitting function is $\mathrm{\psi_{tot}}=\mathrm{\psi_{tot,0}}\left(\frac{\mathrm{E}}{3 \mathrm{EeV}}\right) ^{\gamma_1}$ and $\mathrm{\psi_E}=\mathrm{\psi_{E,0}}\left(\frac{\mathrm{E}}{10 \mathrm{EeV}}\right) ^{\gamma_2}$, respectively.      The corresponding error are calculated by $\rm \sigma_{\mathrm{\psi_{tot}}}=\psi_{tot} \sqrt{ \left(\frac{\sigma_{\mathrm{\psi_{tot,0}}}}{\mathrm{\psi_{tot,0}}}\right)^2+\left(\sigma_{\gamma_1} \ln \frac{\mathrm{E}}{3 \mathrm{EeV}}\right)^2}$ and $\rm \sigma_{\mathrm{\psi_{E}}}=\psi_{E} \sqrt{ \left(\frac{\sigma_{\mathrm{\psi_{E,0}}}}{\mathrm{\psi_{E,0}}}\right)^2+\left(\sigma_{\gamma_2} \ln \frac{\mathrm{E}}{10 \mathrm{EeV}}\right)^2}$.
The proportion and error of the extragalactic cosmic ray flux to the total cosmic ray flux at the dip structure are $\eta=\frac{\psi_E}{\psi_{t o t}}$ and $\sigma_\eta=\eta \sqrt{\left(\frac{\sigma_{\psi_E}}{\psi_E}\right)^2+\left(\frac{\sigma_{\psi_{t o t}}}{\psi_{t o t}}\right)^2}$, respectively.

Finally, using Eq \ref{EqA.5}, we calculated the amplitudes of GCR anisotropy at the dip structure, denoted as $\rm \delta_{G,3dp}$ and $\rm \delta_{G,4dp}$, as presented in Table \ref{tab3}. Their uncertainties are determined through the error propagation formula:
\begin{equation}
\sigma_{\delta_G}=\frac{1}{1-\eta} \sqrt{\sigma_{\delta_{\text {tot }}}^2+\eta^2 \sigma_{\delta_E}^2+\left(\frac{\left(\delta_{\text {tot }}+\delta_E\right) \sigma_{\eta}}{1-\eta}\right)^2}\tag{B.1}
\label{EqB.1}
\end{equation}

\begin{table*}[htbp]
\footnotesize
\centering
\caption{Extrapolation of the extragalactic anisotropy amplitude measured by AUGER at the dip structure. $\rm \delta_{E, 3dp}$ and $\rm \delta_{E, 4dp}$ represent the anisotropy amplitudes obtained from fitting the three highest energy points and all four energy points, respectively. $\rm \delta_{tot}$ is the raw data of anisotropy amplitude observed by AUGER. In the most conservative scenario (MCS), $\rm \delta_{E, MCS}$ gives the upper limits of the extragalactic anisotropy amplitude measured by AUGER with energies below $\rm \sim~5~EeV$.}
\label{tab1}
\begin{tabular}{|c|c|c|c|c|c|c|c|c|c|}
\hline Experiment & AUGERSD750 & \multicolumn{8}{|c|}{AUGERSD1500} \\
\hline E[EeV] & 0.43 & 0.43 & 0.70 & 1.28 & 2.48 & 5.1 & 10.3 & 20.3 & 40 \\
\hline $\rm \delta_{tot}[\%]$ & $0.5_{-0.2}^{+0.6}$ & $0.5_{-0.2}^{+0.5}$ & $0.4_{-0.2}^{+0.3}$ & $0.1_{-0.1}^{+0.4}$ & $0.4_{-0.2}^{+0.3}$ & $1.0_{-0.4}^{+0.6}$ & $5.7_{-0.9}^{+1.0}$ & $5.9_{-1.8}^{+2.0}$ & $11^{+4}_{-3}$ \\
\hline $\rm \delta_{E, 3dp}[\%]$ & $1.43 \pm 1.17$ & $1.43 \pm 1.17$ & 1.76 $\pm 1.22$ & 2.27 $\pm 1.24$ & 3.01 $\pm 1.18$ & - & \multicolumn{3}{|c|}{Fitted} \\
\hline $\rm \delta_{E, 4dp}[\%]$ & $0.20 \pm 0.11$ & $0.20 \pm 0.11$ & 0.32 $\pm 0.15$ & 0.55 $\pm 0.21$ & 1.01 $\pm 0.29$ & \multicolumn{4}{|c|}{Fitted} \\
\hline $\rm \delta_{E, MCS}[\%]$ & $1.0_{-0.4}^{+0.6}$ & $1.0_{-0.4}^{+0.6}$ & $1.0_{-0.4}^{+0.6}$ & $1.0_{-0.4}^{+0.6}$ & $1.0_{-0.4}^{+0.6}$ & \multicolumn{4}{|c|}{Upper limits of EGCRs}\\
\hline
\end{tabular}
\end{table*}

\begin{table*}[htbp]
\centering
\caption{Extrapolation of the extragalactic CR flux measured by AUGER at the dip structure.}
\label{tab2}

\begin{tabular}{|l|l|l|l|l|l|l|l|l|}
\hline E[EeV] & 0.43 & 0.70 & 1.28 & 2.48 & \multicolumn{4}{|c|}{AUGER energy spectrum} \\
\hline lg(E/eV) & 17.63 & 17.85 & 18.11 & 18.39 & 18.45 & 18.55 & 18.65 & - \\
\hline $\rm \psi_{\text {tot}}$ & $905.26^{+24.96}_{-24.96}$& $183.99^{+3.82}_{-3.82}$& $25.57^{+0.32}_{-0.32}$&
$2.941^{+0.011}_{-0.011}$& $1.9383_{-0.0067}^{+0.0067}$ & $0.9076{ }_{-0.0041}^{+0.0042}$ & $0.4310_{-0.0025}^{+0.0025}$ & - \\
\hline $\lg (\mathrm{E} / \mathrm{eV})$ & 17.63 & 17.85 & 18.11 & 18.39 & 18.75 & 18.85 & 18.95 & 19.05 \\
\hline $\psi_{\mathrm{E}}$& $139.01^{+8.43}_{-8.43}$ & $40.85^{+2.10}_{-2.10}$ & $8.96^{+0.36}_{-0.36}$ & $1.700^{+0.047}_{-0.047}$& $0.2164_{-0.0016}^{+0.0016}$ & $0.1227_{-0.0011}^{+0.0011}$ & $0.06582_{-0.0073}^{+0.0074}$ & $0.03796_{-0.00049}^{+0.00049}$ \\
\hline $\eta$ [\%] & $15.4^{+1.0}_{-1.0}$ & $22.2^{+1.2}_{-1.2}$ & $35.0^{+1.5}_{-1.5}$ & $57.8^{+1.6}_{-1.6}$ & \multicolumn{4}{|c|}{-} \\
\hline
\end{tabular}

\begin{tablenotes}
    \footnotesize
    \item[1] The flux $\psi$ is in unit of $\mathrm{km}^{-2} \mathrm{yr}^{-1} \mathrm{sr}^{-1} \mathrm{EeV}^{-1}$ in the table.
\end{tablenotes}
\end{table*}

\begin{table*}[htbp]
\footnotesize
\centering
\caption{Calculated results of galactic anisotropy amplitude at the AUGER dip structure.}
\label{tab3}
\begin{tabular}{|c|c|c|c|c|c|}
\hline Experiment & AUGERSD750 & \multicolumn{4}{|c|}{AUGERSD1500} \\
\hline E[EeV] & 0.43 & 0.43 & 0.70 & 1.28 & 2.48 \\
\hline $\delta_{\text {tot }}$ [\%] & $0.5_{-0.2}^{+0.6}$ & $0.5_{-0.2}^{+0.5}$ & $0.4{ }_{-0.2}^{+0.3}$ & $0.1{ }_{-0.1}^{+0.4}$ & $0.4_{-0.2}^{+0.3}$ \\
\hline $\delta_{\mathrm{E}, 3 \mathrm{dp}}[\%]$ & $1.43 \pm 1.17$ & $1.43 \pm 1.17$ & $1.76 \pm 1.22$ & $2.27 \pm 1.24$ & $3.01 \pm 1.18$ \\
\hline $\delta_{\mathrm{E}, 4 \mathrm{dp}}[\%]$ & $0.20 \pm 0.11$ & $0.20 \pm 0.11$ & $0.32 \pm 0.15$ & $0.55 \pm 0.21$ & $1.01 \pm 0.29$ \\
\hline $\rm \delta_{E, MCS}[\%]$ & $1.0_{-0.4}^{+0.6}$ & $1.0_{-0.4}^{+0.6}$ & $1.0_{-0.4}^{+0.6}$ & $1.0_{-0.4}^{+0.6}$ & $1.0_{-0.4}^{+0.6}$ \\
\hline $\delta_{\mathrm{G}, 3 \mathrm{dp}}[\%]$ & $0.85 \pm 0.52$ & $0.85 \pm 0.46$ & $1.01 \pm 0.47$ & $1.37 \pm 0.77$ & $5.0 \pm 1.7$ \\
\hline $\delta_{\mathrm{G}, 4 \mathrm{dp}}[\%]$ & $0.63 \pm 0.47$ & $0.63 \pm 0.41$ & $0.60 \pm 0.32$ & $0.45 \pm 0.40$ & $2.32 \pm 0.71$ \\
\hline $\rm \delta_{G, MCS}[\%]$ & $2.00 \pm 0.94$ & $2.00 \pm 0.86$ & $1.80 \pm 0.71$ & $1.20 \pm 0.71$ & $1.80 \pm 0.71$ \\
\hline
\end{tabular}
\end{table*}

\section{Kolmogorov test for the distribution of anisotropy amplitudes above the knee structure}\label{appendixC}
Above the knee structure, the anisotropy amplitude distribution predicted by the spectral parameters for GCRs significantly exceeds the experimental observations from AUGER. Therefore, we employ the Kolmogorov test to quantitatively assess whether the two distributions are consistent.

The testing procedure is as follows:

1. We treat the residuals between the experimentally observed anisotropy amplitude $\delta_{\exp}$ and the theoretically predicted anisotropy amplitude $\delta_{\text {cal}}$ at each energy point $E_i$ as random variables. These residuals are standardized to yield random variables that should follow a standard normal distribution:
\begin{equation}
Z\left(E_i\right)=\frac{\delta_{\exp}\left(E_i\right)-\delta_{\text {cal}}\left(E_i\right)}{\sigma\left(E_i\right)}\tag{C.1}
\end{equation}

\begin{equation}
\sigma\left(E_i\right)=\sqrt{\sigma_{\exp}^2\left(E_i\right)+\sigma_{\text {cal}}^2\left(E_i\right)}\tag{C.2}
\end{equation}

2. Null hypothesis: The empirical distribution function of $\rm Z(E_i)$ is consistent with the cumulative distribution function of the standard normal distribution.
Alternative hypothesis: The two distributions are inconsistent.

3. We sort the obtained $Z(E_i)$ values, then compute the empirical distribution function $S_{12}(Z(E_i))$ and the cumulative distribution function $\Phi(Z(E_i))$ of the standard normal distribution at these energy points, thus obtaining the Kolmogorov test statistic $D_{12}=\max \left|S_{12}\left(Z\left(E_i\right)\right)-\Phi\left(Z\left(E_i\right)\right)\right|$.

4. At a given confidence level, the computational results are tested by consulting the critical values table and comparing them accordingly.

When using the modified I in AUGER, the random variable Z for the Kolmogorov test of the anisotropic amplitude distribution above the knee structure is shown in Table \ref{tab4}. The empirical distribution function and the cumulative distribution function of the standard normal distribution are illustrated in the upper panel of Figure \ref{fig7}, where the maximum difference obtained is $\rm D_{12}^{I}=0.412$. By consulting the table, we find that the critical value of the test statistic at a degree of freedom of 12 and $\rm \alpha=0.05$ is 0.375, and $\rm D_{12}^{I}=0.412>0.375$. Similarly, when using the modified II in AUGER, we obtain Table \ref{tab5} and the middle panel of Figure \ref{fig7}, where the maximum difference is $\rm D_{12}^{II}=0.418>0.375$. Therefore, when applying both modifications in AUGER, the results of the Kolmogorov test above the knee structure reject the null hypothesis at the 95$\%$ confidence level.

\begin{table}[htbp]
\footnotesize
\centering
\caption{The calculation results of AUGER using correction I. Only the last five experimental data points in the table have been corrected, while the others remain unadjusted (hereafter the same).}
\label{tab4}
\begin{tabular}{|l|l|l|l|l|l|}
\hline $\mathrm{E}[\mathrm{PeV}]$ & $\delta_{\exp }[\%]$ & $\sigma_{\exp }[\%]$ & $\delta_{\text {cal }}[\%]$ & $\sigma_{\text {cal }}[\%]$ & Z \\
\hline 2.7 & 0.26 & 0.10 & 0.21468 & 0.00027 & 0.45319835 \\
\hline 5.3 & 0.18 & 0.07 & 0.303 & 0.019 & -0.06469295 \\
\hline 6.1 & 0.29 & 0.16 & 0.328 & 0.024 & -0.23487238 \\
\hline 33 & 1.2 & 0.9 & 1.08 & 0.23 & 0.1291817 \\
\hline 51 & 0.8 & 0.4 & 1.47 & 0.35 & -1.26056596 \\
\hline 88 & 0.6 & 0.3 & 2.28 & 0.67 & -2.28852252 \\
\hline 161 & 0.4 & 0.4 & 3.77 & 1.32 & -2.44331231 \\
\hline 430 & 0.85 & 0.52 & 11.38 & 6.96 & -1.508722606 \\
\hline 430 & 0.85 & 0.46 & 11.38 & 6.96 & -1.50963747 \\
\hline 700 & 1.01 & 0.47 & 20.27 & 15.74 & -1.2230889 \\
\hline 1280 & 1.37 & 0.77 & 41.35 & 42.10 & -0.94948491 \\
\hline 2480 & 5 & 1.7 & 90.54 & 120.01 & -0.00712774 \\
\hline
\end{tabular}
\end{table}

Furthermore, if the GCR anisotropy amplitude observed by AUGER reaches its upper limit, can the calculated confidence level in this most onservative scenario (MCS) still reach 95\%? From the phase distribution in Figure \ref{fig3}, the anisotropy phase observed by AUGER points toward an extragalactic direction at energies above 5 EeV, where the anisotropy is dominated by EGCRs (i.e., $\rm \eta= \psi_{E}/\psi_{tot} > 0.5$). At lower energies, the phase shifts from extragalactic to the Galactic direction, and the anisotropy is dominated by GCRs (i.e., $\eta<0.5$). From the amplitude distribution in Figure \ref{fig4}, the anisotropies of EGCRs and GCRs already cancel each other around $\rm 5~EeV$, which provides an upper limit on the anisotropy amplitude of EGCRs below $\rm 5~EeV$ ($\rm \delta_{E,MCS}$ in Table \ref{tab1}). According to Eq \ref{EqA.5}, when $\eta$ and the extragalactic anisotropy amplitude $\rm \delta_E$ increase, the Galactic anisotropy amplitude $\rm \delta_G$ increases. Substituting $\eta = 0.5$ and $\rm \delta_{E,MCS}$ (Table \ref{tab1}) into Equations \ref{EqA.5} and \ref{EqB.1} (where $\eta$ is fixed and $\sigma_{\eta} =0$), we obtain the upper limits on the GCR anisotropy amplitude $\rm \delta_{G,MCS}$ and uncertainties (Table \ref{tab3}). As a result, we obtain Table \ref{tab6} and the lower panel of Figure \ref{fig7}, where the maximum difference is $\rm D_{12}^{MCS}=0.413>0.375$. Consequently, the result of the Kolmogorov test above the knee structure at this most onservative scenario (MCS) also reject the null hypothesis at the 95$\%$ confidence level.

\begin{table}[htbp]
\footnotesize
\centering
\caption{The calculation results of AUGER using correction II.}
\label{tab5}
\begin{tabular}{|l|l|l|l|l|l|}
\hline $\mathrm{E}[\mathrm{PeV}]$ & $\delta_{\exp }[\%]$ & $\sigma_{\exp }[\%]$ & $\delta_{\text {cal }}$ [\%] & $\sigma_{\text {cal }}[\%]$ & Z \\
\hline 2.7 & 0.26 & 0.10 & 0.21468 & 0.00027 & 0.45319835 \\
\hline 5.3 & 0.18 & 0.07 & 0.303 & 0.019 & -0.06469295 \\
\hline 6.1 & 0.29 & 0.16 & 0.328 & 0.024 & -0.23487238 \\
\hline 33 & 1.2 & 0.9 & 1.08 & 0.23 & 0.1291817 \\
\hline 51 & 0.8 & 0.4 & 1.47 & 0.35 & -1.26056596 \\
\hline 88 & 0.6 & 0.3 & 2.28 & 0.67 & -2.28852252 \\
\hline 161 & 0.4 & 0.4 & 3.77 & 1.32 & -2.44331231 \\
\hline 430 & 0.63 & 0.47 & 11.38 & 6.96 & -1.54103058 \\
\hline 430 & 0.63 & 0.41 & 11.38 & 6.96 & -1.54186729 \\
\hline 700 & 0.60 & 0.32 & 20.27 & 15.74 & -1.2230889 \\
\hline 1280 & 0.45 & 0.40 & 41.35 & 42.10 & -0.94948491 \\
\hline 2480 & 2.32 & 0.71 & 90.54 & 120.01 & -0.00712774 \\
\hline
\end{tabular}
\end{table}

\begin{table}[htbp]
\footnotesize
\centering
\caption{The calculation results of AUGER using MCS.}
\label{tab6}
\begin{tabular}{|l|l|l|l|l|l|}
\hline $\mathrm{E}[\mathrm{PeV}]$ & $\delta_{\exp }[\%]$ & $\sigma_{\exp }[\%]$ & $\delta_{\text {cal }}$ [\%] & $\sigma_{\text {cal }}[\%]$ & Z \\
\hline 2.7 & 0.26 & 0.10 & 0.21468 & 0.00027 & 0.45319835 \\
\hline 5.3 & 0.18 & 0.07 & 0.303 & 0.019 & -0.06469295 \\
\hline 6.1 & 0.29 & 0.16 & 0.328 & 0.024 & -0.23487238 \\
\hline 33 & 1.2 & 0.9 & 1.08 & 0.23 & 0.1291817 \\
\hline 51 & 0.8 & 0.4 & 1.47 & 0.35 & -1.26056596 \\
\hline 88 & 0.6 & 0.3 & 2.28 & 0.67 & -2.28852252 \\
\hline 161 & 0.4 & 0.4 & 3.77 & 1.32 & -2.44331231 \\
\hline 430 & 2.00 & 0.94 & 11.38 & 6.96 & -1.3355754 \\
\hline 430 & 2.00 & 0.86 & 11.38 & 6.96 & -1.33752922 \\
\hline 700 & 1.80 & 0.71 & 20.27 & 15.74 & -1.17225145 \\
\hline 1280 & 1.20 & 0.71 & 41.35 & 42.10 & -0.95354612 \\
\hline 2480 & 1.80 & 0.71 & 90.54 & 120.01 & -0.00739438 \\
\hline
\end{tabular}
\end{table}

\end{appendices}

\bibliography{sample631}{}
\bibliographystyle{aasjournal}



\end{document}